\documentclass[conference]{IEEEtran}
\IEEEoverridecommandlockouts
\usepackage{cite}
\usepackage{amsmath,amssymb,amsfonts}
\usepackage{algorithmic}
\usepackage{enumitem}
\usepackage{graphicx}
\usepackage{textcomp}
\usepackage{xcolor}
\usepackage{tikz}
\usetikzlibrary{positioning, arrows.meta, shapes.geometric, calc}
\def\BibTeX{{\rm B\kern-.05em{\sc i\kern-.025em b}\kern-.08em
    T\kern-.1667em\lower.7ex\hbox{E}\kern-.125emX}}
\usepackage{eso-pic}

\begin{document}
\title{\vspace{1.5cm}Blockchain-Enabled Explainable AI for Trusted Healthcare Systems}

\author{\IEEEauthorblockN{Md Talha Mohsin}
\IEEEauthorblockA{\textit{Department of Finance \& Operations Management} \\
\textit{University of Tulsa} \\
Tulsa, OK 74104, USA \\
}
}

\maketitle

\begin{abstract}
This paper introduces a Blockchain-Integrated Explainable AI Framework (BXHF) for healthcare systems to tackle two essential challenges confronting health information networks: safe data exchange and comprehensible AI-driven clinical decision-making. Our architecture incorporates blockchain, ensuring patient records are immutable, auditable, and tamper-proof, alongside Explainable AI (XAI) methodologies that yield transparent and clinically relevant model predictions. By incorporating security assurances and interpretability requirements into a unified optimization pipeline, BXHF ensures both data-level trust (by verified and encrypted record sharing) and decision-level trust (with auditable and clinically aligned explanations). Its hybrid edge-cloud architecture allows for federated computation across different institutions, enabling collaborative analytics while protecting patient privacy. We demonstrate the framework's applicability through use cases such as cross-border clinical research networks, uncommon illness detection and high-risk intervention decision support. By ensuring transparency, auditability, and regulatory compliance, BXHF improves the credibility, uptake, and effectiveness of AI in healthcare, laying the groundwork for safer and more reliable clinical decision-making.
\end{abstract}


\section{Introduction}

Artificial intelligence (AI) has become ubiquitous in a multitude of industries, including finance, healthcare, retail, and communications. In healthcare, where user needs have paramount importance, AI has been powering various applications. These include disease diagnosis and prognosis, personalized treatment recommendations, drug development, increasing operational efficiency and more. Using large-scale patient records, medical imaging, and genomic data, AI have shown great capability in identifying subtle patterns that often remain undetected by human experts. However, the widespread adoption of AI in healthcare continues to face two persistent challenges: the lack of interpretability in AI-driven predictions and the integrity of medical data sharing across institutions. Healthcare institutions rely heavily on cross-institutional data interchange to increase diagnostic accuracy and accelerate medical research. However, concerns about data manipulation, unauthorized access, and a lack of trust among stakeholders impede effective collaboration. At the same time, many cutting-edge AI algorithms operate as "black boxes", making correct predictions without providing meaningful explanations. This lack of transparency often undermines clinician trust and user friendliness and raises ethical concerns about accountable clinical decision-making. In addition, data Integrity is also one of the most grave concern for the current healthcare industry  {\cite{zarour_data_integrity_2021}. By handling sensitive healthcare data in a decentralized, securely, and transparent manner, blockchain solves data security, interoperability, and patient data ownership issues \cite{mazhar_generative_2025}.\\

Blockchain technology has emerged as a potential solution for the secure exchange of information and the preservation of data integrity. Introduced in the cryptocurrency Bitcoin in 2008, it is a chain of time-stamped blocks which are linked using cryptographic hashes  \cite{holbl_blockchain_2018} and has inherited characteristics such as decentralization, transparency and anonymization \cite{nakamoto_bitcoin_2008}.  The patient centric mechanisms of the healthcare industry make it apt for blockchain \cite{mettler_blockchain_2016} and it could be a technology that may potentially help in personalized, reliable, and secure healthcare, \cite{siyal_blockchain_2019}.  Also, as the infrastructures of healthcare organizations consist of connected devices and software applications which communicate with other IT systems, they are also heavily affected by the use of blockchain and the Internet of Things (IoT) \cite{farouk_blockchain_2020}.  Some of the challenges healthcare systems face could be massively improved by incorporating blockchain and Explainable AI (XAI) in their systems. The challenges include pharmaceutical product integrity, clinical trial efficiency, fraud, advancing personalized medicine \cite{omidian_synergizing_2024}, fragmented and slow access to medical data, system compatibility, and  data quality \cite{medrec_2018}.

To solve these issues, we propose a Blockchain-Integrated Explainable AI (XAI) Framework for healthcare data exchange. Blockchain, as a a decentralized, distributed, and immutable digital ledger, allows the secure, tamper-proof, and auditable interchange of medical records between institutions, supporting Clinical Decision Support (CDS) systems. XAI, on the other hand, make AI-driven recommendations interpretable and clinically validated. By combining these two paradigms, the framework not only strengthens data integrity but also builds confidence in AI-supported diagnoses and treatments.

\section{Background and Related Work}

\subsection{Blockchain in Healthcare}

Blockchain is a verified decentralized and public digital ledger which records transactions on many networks using cryptography  so that no record involved can be altered retroactively without altering any blocks afterwards  \cite{prokofieva_blockchain_2019} \cite{haleem_blockchain_2021}. A key attribute of blockchain is decentralization; as no central authority controls the content added to the blockchain, the entries passed on to the blockchain are agreed upon in a peer-to-peer network \cite{hasselgren_blockchain_2020}.
Its decentralized ledger facilitates tamper-proof record keeping, while immutability, traceability, and smart contracts improve transparency and control over access permissions. \\
Because of the need for a patient-centric approach to healthcare systems and to connect disparate systems and increase the accuracy of electronic healthcare records (EHRs),  blockchain has tremendous potential is healthcare \cite{holbl_blockchain_2018} and blockchain applications in healthcare have been the subject of numerous studies. The decentralized nature, openness and permissionless of blockchain offer a unique solution for healthcare \cite{prokofieva_blockchain_2019}, as both data sharing and access is an inherent problem with civilian health records \cite{roehrs_omniphr_2017}. Some of the key properties of blockchain, such as immutability, decentralization, and transparency, can potentially address pressing issues in healthcare, including access to patients’ own health information and incomplete records at point of care \cite{zhang_blockchain_2018}. 
Blockchain can also be used to circumvent problems in traditional healthcare architecture for secure storage, sharing and retrieval of EHRs \cite{jayabalan_scalable_2022}.  Blockchain being s a distributed architecture with decentralized and tamper-proof features can provide a new, better way to protect the personal health records sharing system \cite{wang_blockchain_2019}. As Privacy and security preservation are imperative when sharing electronic health records \cite{farouk_blockchain_2020}, the integration of Blockchain in healthcare Technology  gives patients greater power over their personal information; thus improving confidentiality and privacy \cite{haleem_blockchain_2021}. Additionally, blockchain will not only provide utmost privacy but also ensure that appropriate users can easily add to and access a permanent record of information \cite{engelhardt_hitching_2017}.

\subsection{Explainable AI in Healthcare}
Along with reliable data interchange, the interpretability of AI models has become crucial for clinical applications. While deep learning models achieve excellent predictive accuracy in medical imaging, genomics, and risk stratification, their "black-box" nature limits clinician trust and creates regulatory concerns. As an AI system should have the capacity being accountable at every stage of the healthcare pathway \cite{procter_holding_2023}, explainable AI (XAI) approaches this problem by providing insights into model reasoning via feature attribution, visualization, rule extraction, and counterfactual explanations.  
Because of the stellar effect of XAI, naturally, there is a large and growing effort to enhance trust and improve acceptance of AI based technology in clinical medicine \cite{hossain_explainable_2025}. One of the  most important reason to incorporate explainability in the sense of scientific or causal explanation in day-to-day clinical practice is the potential for improving future care by building a more robust model of the world \cite{pierce_explainability_2022}.

Despite individual developments in blockchain and XAI, little efforts have been made to merge the two paradigms into a cohesive framework. Existing research focuses on either secure, decentralized data sharing or the interpretability of AI-driven predictions, but not both. This division hinders the possibility of developing truly trustworthy healthcare systems, in which clinicians and institutions can rely on both the integrity of shared data and the transparency of AI-assisted decision-making.

\section{Proposed Framework}

We propose a unified \textbf{Blockchain--XAI Healthcare Framework (BXHF)} that simultaneously ensures secure data provenance and interpretable clinical predictions in supporting Clinical Decision Support (CDS) within healthcare systems. Unlike prior works that treat blockchain-based healthcare security and explainable AI as isolated research directions, BXHF establishes a mathematically grounded pipeline in which security guarantees and interpretability constraints are jointly encoded into the predictive process.

\subsection{System Architecture}

The BXHF system is composed of five interdependent layers, as illustrated below in Fig \ref{fig:1}. 

\begin{figure}[htbp]
    \centering
    \resizebox{\columnwidth}{!}{%
    \begin{tikzpicture}[node distance=1.2cm]

    \tikzset{
        layer/.style={rectangle, rounded corners=6pt, draw=black!70, thick,
                      minimum width=2.4cm, minimum height=0.8cm, align=center, fill=blue!10},
        process/.style={rectangle, draw=black!80, thick, fill=green!10,
                      minimum width=2.6cm, minimum height=0.9cm, align=center},
        db/.style={cylinder, shape border rotate=90, draw, thick,
                   minimum height=1cm, minimum width=0.8cm, aspect=0.25, fill=yellow!20},
        cloud/.style={ellipse, draw=black!70, thick, fill=gray!20, minimum width=2cm, align=center},
        arrow/.style={-{Latex[length=2.5mm]}, thick},
    }

    \node[layer] (hosp1) {Hospital A \\ (Edge Node)};
    \node[layer, right=2.5cm of hosp1] (hosp2) {Hospital B \\ (Edge Node)};

    \node[process, below=1.5cm of $(hosp1)!0.5!(hosp2)$] (blockchain) 
        {Blockchain Layer \\ (Distributed Ledger + Smart Contracts)};

    \node[db, below left=1.2cm and 0.3cm of blockchain] (edb1) {Encrypted DB A};
    \node[db, below right=1.2cm and 0.3cm of blockchain] (edb2) {Encrypted DB B};

    \node[process, below=2.3cm of blockchain] (ai) 
        {AI Layer \\ Predictive Models};

    \node[process, below=1.2cm of ai, fill=orange!20] (xai) 
        {XAI Layer \\ SHAP / LIME / Attention Maps};

    \node[cloud, below=1.8cm of xai] (deploy) {Edge–Cloud Deployment \\ (Federated Access)};

    \draw[arrow] (hosp1.south) -- ++(0,-0.6) -| (blockchain.north);
    \draw[arrow] (hosp2.south) -- ++(0,-0.6) -| (blockchain.north);

    \draw[arrow] (blockchain.south) -- (ai.north);
    \draw[arrow] (ai.south) -- (xai.north);
    \draw[arrow] (xai.south) -- (deploy.north);

    \draw[arrow] (blockchain.south west) -- (edb1.north);
    \draw[arrow] (blockchain.south east) -- (edb2.north);

    \draw[dashed, arrow] (hosp1.east) -- (hosp2.west) node[midway, above, sloped] {\tiny Federated Sync};

    \end{tikzpicture}%
    }
    \caption{ Workflow architecture of the Blockchain–XAI Healthcare Framework (BXHF).}
    \label{fig:1}
\end{figure}
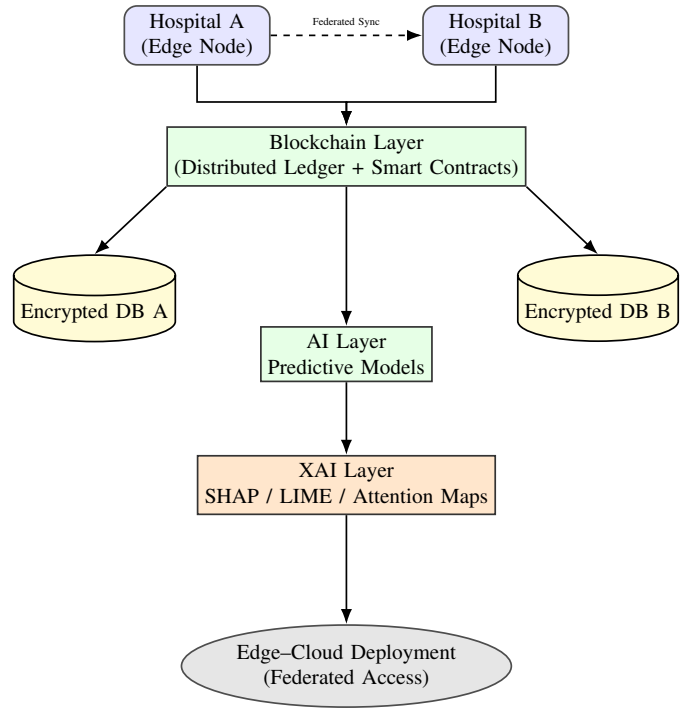

\subsubsection{Data Layer (Encrypted Patient Records)}
Patient records are defined as:
\[
D = \{(x_i, y_i)\}_{i=1}^n
\]
where $x_i$ denotes multimodal medical features (EHRs, imaging, lab results) and $y_i$ denotes clinical outcomes. Each $x_i$ is encrypted using homomorphic encryption. As raw data access is disallowed in the framework, computations are performed under privacy-preserving query protocols, ensuring confidentiality and compliance with regulations such as HIPAA and GDPR.

\subsubsection{Blockchain Layer (Immutable Audit and Access Control)}
A distributed ledger $L$ maintains a hash-based mapping between encrypted data identifiers and access transactions. Smart contracts $\phi$ are deployed to enforce access policies:
\[
\phi(u, d) =
\begin{cases} 
1, & \text{if user } u \text{ is authorized to access data } d \\
0, & \text{otherwise.}
\end{cases}
\]
Each of the read/write operation is permanently auditable, establishing accountability and data provenance across stakeholders.

\subsubsection{AI Layer (Predictive Modeling)}
Let $f: X \to Y$ be the predictive model trained under secure data retrieval. For a given patient input $x$, the prediction is:
\[
\hat{y} = f(x), \quad f^* = \arg\min_{f \in \mathcal{F}} \; \mathbb{E}_{(x,y)\sim D} \; [\ell(f(x), y)]
\]
where $\ell$ is the task-specific loss function (e.g., cross-entropy for classification or mean squared error for prognosis). Model training leverages federated nodes, ensuring that no raw data leaves institutional boundaries.

\subsubsection{XAI Layer (Formalized Interpretability)}
An explanation function $g$ provides interpretable insights:
\[
g: (x, f(x)) \mapsto E
\]
where $E$ denotes human-understandable representations (e.g., feature attribution scores $\alpha_j$, saliency maps, or rule-based sets). Feature importance must satisfy:
\[
\sum_j \alpha_j = f(x).
\]

Unlike post-hoc methods, BXHF introduces \emph{constraint-augmented training}:
\[
f^* = \arg\min_{f \in \mathcal{F}} \; \mathbb{E}\big[\ell(f(x), y)\big] + \lambda \cdot \Omega(g(f(x)))
\]
where $\Omega$ penalizes explanations that lack clinical plausibility. This ensures interpretability is enforced during optimization rather than appended afterward.

\subsubsection{Deployment Layer (Hybrid Edge--Cloud)}
Sensitive computations (encryption, preliminary inference) are executed on hospital-owned edge devices. Large-scale training is handled by federated cloud nodes. The blockchain guarantees trust and consistency across nodes, eliminating the need for a central authority.

\subsubsection{Integrated Trust Guarantee (Security + Interpretability)}

The BXHF framework integrates blockchain security and explainable AI (XAI) into a single, mathematically grounded objective. 
Rather than treating these aspects separately, the framework ensures that predictive accuracy, interpretability, and data security are optimized together. 

{\small
\[
\mathcal{J}(f, D) = \min_{f \in \mathcal{F}} 
\underbrace{\mathbb{E}_{(x,y)\sim D}[\ell(f(x), y)]}_{\text{Loss}}
+ \underbrace{\lambda_1 \Omega(g(f(x)))}_{\text{Interp}}
- \underbrace{\lambda_2 \mathcal{S}(D)}_{\text{Sec}},
\]
}

where:
\begin{itemize}
    \item $f \in \mathcal{F}$ is the predictive model from the set of allowable models $\mathcal{F}$,
    \item $x$ represents the patient input features (EHR, imaging, lab results),
    \item $y$ represents the corresponding clinical outcome,
    \item $\ell(f(x), y)$ is the task-specific predictive loss function, such as cross-entropy for classification or mean squared error for regression,
    \item $g(f(x))$ is the explanation function that provides interpretable insights for a prediction,
    \item $\Omega(g(f(x)))$ penalizes explanations that are inconsistent or lack clinical plausibility,
    \item $\mathcal{S}(D)$ is a security measure that quantifies the integrity, provenance, and auditability of the data $D$ stored on the blockchain,
    \item $\lambda_1 > 0$ and $\lambda_2 > 0$ are trade-off hyperparameters controlling the balance between interpretability, security, and predictive performance.
\end{itemize}

This formulation creates a dual-layer trust mechanism:  
\begin{enumerate}[label=\roman*.]
    \item \textbf{Data-level trust:} Blockchain ensures that all patient data is immutable, traceable, and auditable. 
          Each access and update is recorded as a hash-based transaction, providing tamper-evident provenance.
    \item \textbf{Decision-level trust:} XAI ensures that predictions are interpretable and explainable. 
          The explanations themselves are cryptographically bound to the blockchain, guaranteeing that they cannot be manipulated after generation.
\end{enumerate}

By jointly optimizing $\mathcal{J}(f, D)$, BXHF ensures that high predictive performance is only achievable if both interpretability and data security constraints are satisfied. 
This mathematically enforces the principle that trustworthy AI requires not only accurate predictions but also transparent reasoning and verifiable data provenance.

\subsection{Workflow Example}

To demonstrate the operational feasibility of the proposed Blockchain-Integrated Explainable Healthcare Framework (BXHF), we describe a workflow that highlights its multi-layered approach to data exchange, model invocation, and explanation integrity. Unlike linear pipelines, BXHF creates an auditable and verifiable process that links predictions and explanations into a tamper-proof ledger. This creates an immutable provenance for medical decision making. \\

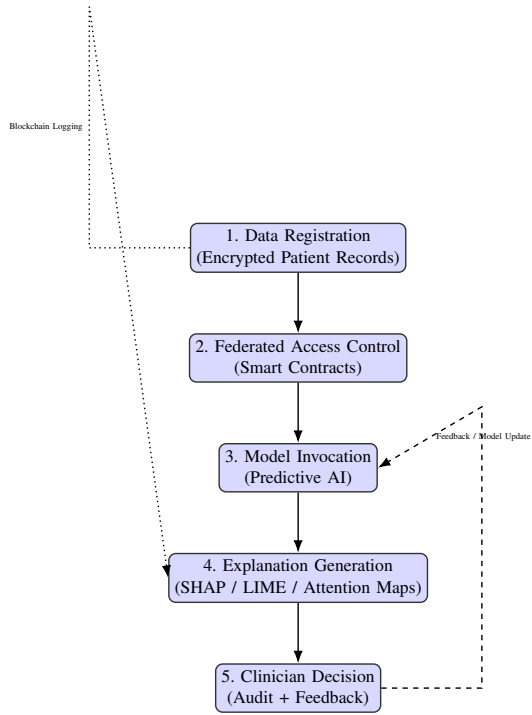
\begin{figure}[htbp]
\centering
\resizebox{0.8\columnwidth}{!}{%
\begin{tikzpicture}[node distance=1.2cm, align=center]

\tikzset{
    stage/.style={rectangle, draw=black!80, thick, fill=blue!15,
                  minimum width=3cm, minimum height=0.9cm, rounded corners=4pt},
    arrow/.style={-{Latex[length=2.5mm]}, thick},
}

\node[stage] (data) {1. Data Registration \\ (Encrypted Patient Records)};
\node[stage, below=of data] (access) {2. Federated Access Control \\ (Smart Contracts)};
\node[stage, below=of access] (model) {3. Model Invocation \\ (Predictive AI)};
\node[stage, below=of model] (explain) {4. Explanation Generation \\ (SHAP / LIME / Attention Maps)};
\node[stage, below=of explain] (clinician) {5. Clinician Decision \\ (Audit + Feedback)};

\draw[arrow] (data.south) -- (access.north);
\draw[arrow] (access.south) -- (model.north);
\draw[arrow] (model.south) -- (explain.north);
\draw[arrow] (explain.south) -- (clinician.north);

\draw[arrow, dashed] (clinician.east) -- ++(2,0) -- ++(0,5.6) -- (model.east) node[midway, right] {\tiny Feedback / Model Update};

\draw[arrow, dotted] (data.west) -- ++(-2,0) -- ++(0,4.8) node[midway, left] {\tiny Blockchain Logging} -- (explain.west);

\end{tikzpicture}%
}
\caption{Use-case flow of BXHF}
\label{fig:2}
\end{figure}

Fig \ref{fig:2} shows secure data registration, federated access, predictive modeling, explainable outputs, and clinician feedback. Blockchain logging ensures auditability at multiple stages.
\textbf{Step 1: Data Registration}\\
With a unique cryptographic hash, all of patient records, including EHRs, medical imaging, and test results, are registered on the blockchain. \\
\textbf{Step 2: Federated Access Control}\\
When an institution or physician wants access to patient information, a smart contract enforces consent-based and regulatory-compliant standards (e.g., HIPAA, GDPR).
Federated access protects sensitive information by limiting sharing to verified entities.\\
\textbf{Step 3: Model Invocation with Provenance}\\
Once data is retrieved, the predictive model is used to make diagnostic or therapeutic suggestions. An explanation vector, such as SHAP values, feature importance scores, or symbolic rule traces, is calculated simultaneously. Predictions and explanations are cryptographically bound and logged as new decision blocks on the blockchain.\\
\textbf{Step 4: Multi-Institutional Validation}\\
Other institutions can independently verify that the model used the correct data inputs and that the explanation is accurately related to the forecast result.
This stage converts explanations from temporary artifacts to auditable entities available across the consortium.\\
\textbf{Step 5: Clinical Interface}\\
Clinicians use a specific interface to get the framework's prediction and interpretable rationale. As each output includes a blockchain-backed integrity certificate, it ensures that the explanation has not been edited after the fact. Also, the provenance chain of the framework ensures trust in both the data and the reasoning behind AI-driven decisions.

For example, let's say a patient is admitted to a hospital with suspected cardiac problems. The lab results as well as the encrypted ECG are uploaded to the system. The clinician then take advantage of the BXHF; the blockchain logs queries, checks access privileges, and provides encrypted data references. The model then predicts heart failure and provides an interpretable explanation: identifying higher troponin levels and aberrant ECG signals as significant drivers. The ledger then records both predictions and explanations, ensuring transparency and auditability.

\subsection{Novelty of BXHF}

BXHF integrates blockchain for healthcare data sharing with explainable AI (XAI) for model transparency, advancing both areas as a whole. BXHF is unique in that it offers dual trust at both the data and decision levels.

\begin{itemize}

\item\textbf {On-chain Explanation Integrity}:
The inherent quality of our architecture ensures that the records include the cryptographic hash of the model-generated explanation, as well as the forecast.  Users can then examine such explanations to ensure they are valid, untampered, and consistent with the model's decision process.
\item\textbf {Two-layered trust mechanism}:
BXHF presents a two-layer trust paradigm:
\begin{itemize}
    \item \textbf{Data-level trust:} The blockchain aspect of the framework provides users with data-level trust through immutability, provenance tracking, and regulated sharing of sensitive healthcare data.
    \item \textbf{Decision-level trust:} The XAI aspect of the modules makes the model predictions interpretable, while BXHF's on-chain storage of explanation proofs ensures that decision rationales can be verified by all stakeholders.
\end{itemize}

\item\textbf {Joint Security-Interpretability Formalism}: BXHF has joint Security-Interpretability Formalism which integrates both security and interpretability into a single optimization pipeline, not as separate concerns.
\item\textbf {Regulatory Alignment Through Design}:
As auditability is weaved into decision-making processes of every model output and explanation, it generates a blockchain trail, ensuring transparent regulatory compliance including HIPAA, GDPR, and FDA guidelines.

\item\textbf {Immutable Explanation Provenance}: As all of the explanations are kept and auditable on blockchain, it makes sure to prohibit any kind of manipulation of post-hoc interpretation outputs.
\end{itemize}

\begin{figure}[htbp]
    \centering
    \includegraphics[width=0.48\textwidth]{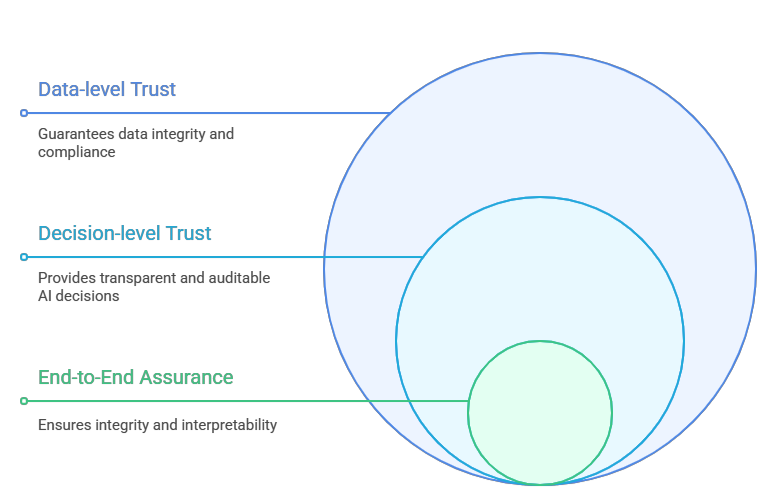} 
    \caption{Two layer Trust}
    \label{fig:3}
\end{figure}

\section{Key Features and Advantages}

Our proposed Blockchain-Integrated Explainable AI Framework (BXHF) has multiple advantages that alleviates the some of the key challenges in healthcare data sharing and AI adoption:

\begin{itemize}

\item\textbf{Secure and Tamper-Proof Data Sharing}\\
Blockchain ensures that all patient data exchanged across institutions is immutable, traceable, and auditable. Each transaction is recorded, which prevents unauthorized modifications and data leakage. 

\item\textbf{Transparent and Interpretable AI Predictions}\\
Instead of giving opaque outputs, the system provides explanations—such as which clinical variables influenced a decision—so physicians can judge whether the recommendation is reasonable.

\item\textbf{Dual-Layer Trust}\\
The framework supports confidence both in the accuracy of the underlying records and in the reasoning of the model itself. Most existing systems address one of these, but BXHF combines the two (Fig \ref{fig:3}).

\item\textbf{Support for compliance}\\
As all of the predictions and their rationales are logged, the framework naturally produces an auditable history that can help with regulations like HIPAA or GDPR. 

\item\textbf{Scalability in deployment}\\
The hybrid edge–cloud architecture allows sensitive computations to occur locally, reducing latency and preserving privacy, while leveraging cloud nodes for large-scale model training and cross-institution collaboration. This design balances efficiency, security, and flexibility.

\item\textbf{Interoperability}\\
By enabling secure exchange of data and model outputs, the framework makes it easier for hospitals or research centers to work together on larger studies or shared clinical trials.

\item\textbf{Clinical Reliability and Patient Safety}\\
Combining verifiable records with interpretable recommendations reduces the risk of errors that could arise from corrupted data or unexplained model outputs.
\end{itemize}

\section{Use Cases}

The Blockchain-Integrated Explainable AI Framework (BXHF) is intended to be adaptable and relevant to a variety of healthcare scenarios requiring data integrity and interpretability. The examples below demonstrate its potential impact:

\begin{enumerate}[label=\roman*.]
\item\textbf{Collaboration between multiple hospitals to diagnose rare diseases}

Patient data from several hospitals is frequently pooled to gain statistically significant insights into rare diseases. BXHF facilitates the secure sharing of sensitive patient records between institutions while also ensuring that AI-generated diagnostic predictions are interpretable. Clinicians can validate the input data and understand why the model recommends a specific diagnosis, which boosts confidence in cross-institutional decision-making.

\item\textbf{Global Clinical Research Collaboration}.
Our framework will allow for research collaboration across borders. As clinical data sharing are frequently subject to varying regulations, the hybrid edge-cloud architecture of BXHF, paired with blockchain, will ensure data provenance and access compliance; XAI will ensure that the AI model outputs are auditable and standardized, which will allow researchers to evaluate results, replicate analyses, and foster trust in the research.

\item\textbf{Federated Learning for Predictive Analytics}

BXHF will make it possible for hospitals to train models together without ever moving patient data outside their own systems. Each hospital keeps full control of its records, while the blockchain layer tracks where updates come from and who has permission to use them. The XAI layer then helps doctors and researchers evaluate the results of these models and the reasoning behind them, thus making the updates clinically useful and trustworthy.

\item\textbf{Provide clinical decision support for high-risk interventions}
In high-risk treatments such as organ transplants, oncology therapy, and intensive care interventions where data integrity and model interpretability are essential, BXHF framework can safely incorporate patient history, imaging, and lab findings, while also simultaneously explaining AI-driven risk estimates and reduce the possibility of errors that could jeopardize patient safety.

\end{enumerate}

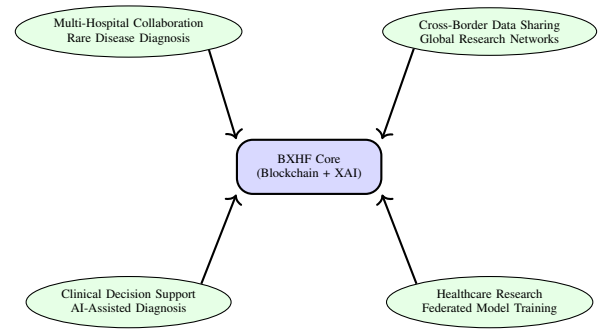
\begin{figure}[htbp]
\centering
\begin{tikzpicture}[every node/.style={font=\footnotesize}, scale=0.6, transform shape]

\node[draw, rectangle, rounded corners=6pt, thick, fill=blue!15, 
      minimum width=3.2cm, minimum height=1.2cm, align=center] 
      (bxhf) at (0,0) {BXHF Core \\ (Blockchain + XAI)};

\node[draw, ellipse, fill=green!10, minimum width=3.2cm, minimum height=1.0cm, 
      align=center] (rare) at (-4,3) {Multi-Hospital Collaboration \\ Rare Disease Diagnosis};

\node[draw, ellipse, fill=green!10, minimum width=3.2cm, minimum height=1.0cm, 
      align=center] (cross) at (4,3) {Cross-Border Data Sharing \\ Global Research Networks};

\node[draw, ellipse, fill=green!10, minimum width=3.2cm, minimum height=1.0cm, 
      align=center] (clinical) at (-4,-3) {Clinical Decision Support \\ AI-Assisted Diagnosis};

\node[draw, ellipse, fill=green!10, minimum width=3.2cm, minimum height=1.0cm, 
      align=center] (research) at (4,-3) {Healthcare Research \\ Federated Model Training};

\draw[->, thick] (rare.south east) -- (bxhf.north west);
\draw[->, thick] (cross.south west) -- (bxhf.north east);
\draw[->, thick] (clinical.north east) -- (bxhf.south west);
\draw[->, thick] (research.north west) -- (bxhf.south east);

\end{tikzpicture}
\caption{Use cases of the BXHF framework demonstrating secure and interpretable AI applications across multiple healthcare scenarios.}
\label{fig:bxhf_usecases_elab}
\end{figure}

\section{Conclusion}

The Blockchain-Integrated Explainable AI Framework (BXHF) tackles two significant challenges present in healthcare: secure data sharing and interpretable AI-driven clinical judgments. BXHF possesses considerable potential to expedient healthcare collaboration and decision-making. As challenges like interoperability issues, user interface complexities, and data security concerns hinder Existing EHR systems \cite{rathore_synergy_2025}, our framework facilitates multi-hospital data exchange, telemedicine, international research partnerships, and critical clinical interventions while ensuring regulatory adherence and patient safety. The framework's versatile design enables its implementation in additional fields necessitating dual trust in data integrity and AI interpretability.

In summary, BXHF illustrates that secure, interpretable, and auditable AI systems are attainable in healthcare, establishing a basis for reliable AI implementation, improved patient outcomes, and strengthened inter-institutional collaboration. The multi-tiered architecture of the framework facilitates strong trust at both the data and decision levels. The disruptive character of blockchain will also strongly affect the balance of power among different market players in healthcare for the better \cite{mettler_blockchain_2016}. Subsequent implementations and pilot studies will further substantiate the framework's practical advantages and scalability in actual clinical settings.

\end{document}